\documentclass[prl,showpacs]{revtex4}

\usepackage{graphicx}
\usepackage{rotating}
\usepackage{amsmath}
\usepackage{amsfonts}
\usepackage{amssymb}
\usepackage{enumerate}
\usepackage{longtable}
\usepackage{dcolumn}
\usepackage{bm}

\newcommand{\be}{\begin{equation}}
\newcommand{\ee}{\end{equation}}
\newcommand{\bn}{\begin{eqnarray}}
\newcommand{\en}{\end{eqnarray}}

\begin{document}

\author {M.S. Laad}
\title {On the Impossibility of Electronic Phase Separation in the $D=1,2$ 
Dimensional Hubbard
Model at $T=0$.
}
\affiliation{Max-Planck-Institut f\"ur Physik Komplexer Systeme,
01187 Dresden, Germany
}

\date{\rm\today}

\begin{abstract}
We show explicitly that the one- and two-dimensional Hubbard model does not show
 phase separation at
any filling away from half-filling at $T=0$.  Apart from a single plausible
assumption, only known exact results and symmetry properties for the one-band
Hubbard model are used.  Implications for usage of the simple one-band
Hubbard model for analysing the physics of cuprate superconductors are
discussed.  We also discuss models where electronic phase separation might generically occur,
in the specific context of colossal magneto-resistance manganites.
\end{abstract}
\pacs{}

\maketitle

  The normal state properties of oxide superconductors are as perplexing as their
high transition temperatures.  Considerable experimental evidence has revealed
that many high $T_{c}$ superconductors have a regime way from half-filling in
which electronic phase separation occurs.  Such a phase occurs between an
oxygen-poor and an oxygen-rich phase that goes superconducting.  It is believed
in some works,
 therefore, that the presence of electronic phase separation (EPS) is
connected with the physics of these doped materials.

  Au {\it et al} and Su [3,4] have obtained rigorous results on phase separation
 in the 2D Hubbard model on a bipartite lattice.  However, in [4], the question
concerning the possibility of phase separation cannot be resolved at T=0.  If
EPS would indeed exist in the $2D$ Hubbard model at $T=0$, one would suppose
the precursor effects associated with this $T=0$ instability to manifest
themselves in the physical response of cuprates, especially near the carrier
doping driven $T=0$ Mott transition.  For example, a singular compressibility
is implied near such a quantum transition; this would necessarily affect the
charge dynamics in a drastic way near the IMT.

  In this paper, we make use of an inequality due to Pitaevskii {\it et al} [5]
 to address this issue.  Unlike in [4], which makes use of rigorous results
on the 2D Hubbard model at $ T>0 $ for a square lattice, this inequality is
valid for T=0, and contains information about the quantum fluctuations in the
system.

  We start with a one-band Hubbard model defined on a square lattice with M
sites in an external space-dependent magnetic field

\be
H=-t\sum_{<ij>,\sigma}(C_{i\sigma}^{\dag}C_{j\sigma}+h.c) +
U\sum_{i}n_{i\uparrow}n_{i\downarrow} 
- \mu\sum_{i\sigma}n_{i\sigma}
\ee
with a staggered external Zeeman magnetic field, $- (h/2)\sum_{i\sigma}\sigma n_
{i\sigma}e^{-i{\bf q}.{\bf R_{i}}}$,
 where the symbols have their usual meanings.  We will consider eqn (1) with bot
h signs of U.  Inspite of intensive study, very few exact results are known for
this model.  It has been shown that there is no spontaneous magnetic order in
one- and two dimensions at any finite temperature, as is expected from the const
raints imposed by the Mermin Wagner theorem [6].
The exact solution of
the 1D Hubbard model has been worked out by Lieb and Wu [7] using the
Bethe ansatz.

  Pitaevskii's inequality, which provides direct information on quantum fluctuations
of the system, reads [5]
\be
\langle[A^{\dag},A]_{+}\rangle\langle[B^{\dag},B]_{+}\rangle \ge |\langle[A^{\dag},B]\rangle|^{2}
\ee
for any two operators A and B satisfying $ \langle0|A|0\rangle $ = 0 = $\langle0
|B|0\rangle $ ($ |0\rangle $ is the ground state of H).
Here, $[A^{\dag},A]_{+}$ denotes the anticommutator of $A$ and $A^{\dag}$
and ${\bf g}$ is fixed by the condition
 
 $e^{i{\bf g\cdot \bf R_{i}}} = 1$ if ${\bf R_{i}}$ connects sites on the same
sublattice, and equals -1 otherwise.
Eqn (2) then gives
\be
S^{\perp}({\bf q} + {\bf g})S^{\perp}({\bf q}) \ge [s_{\bf g}^{z}]^2
\ee
where we used $[S^{+},S^{-}]=2s^{z}$ and introduced the static structure factor
$ S^{\perp}({\bf k})=(1/2N)\langle[S_{+}^{\dag}({\bf k}),S_{+}({\bf
k})]_{+}\rangle $ 
along with the staggered order parameter
\be
 s_{\bf g}^{z} = (1/N)\langle\Sigma_{i}s_{i}^{z}e^{i{\bf g\cdot R_{i}}}\rangle
\ee
 for the antiferromagnet.
Following [5], we consider the Cauchy-Schwarz inequality,
\be
S^{\perp}({\bf q}) = \int{S^{\perp}({\bf q},\omega) d\omega}
                    \le \sqrt{\int{\omega S^{\perp}({\bf q},\omega) d\omega} \int{\frac{S^{\perp}({\bf q},\omega)}{\omega} d\omega}}
\ee
and introduce an effective speed of spin excitations by
\be
c_{eff}^{2}({\bf q})q^{2} = {\int{\omega S^{\perp}({\bf q},\omega) d\omega}\over
{\int{\frac{S^{\perp}({\bf q},\omega)}{\omega} d\omega}}}
\ee
so that eqn (3) becomes
\be
 S^{\perp}({\bf q}+{\bf g}) \ge {{s_{\bf g}^{2}} \over {qc_{eff}({\bf q})\chi^{\perp}({\bf q})}}
\ee
where $ \chi^{\perp}({\bf q}) = \int{{S^{\perp}({\bf q},\omega) \over \omega}d\omega} $ 
is the transverse static response function.

To proceed, we consider
\be
\int{\omega S^{\perp}({\bf q}, \omega) d\omega} = (1/2N)\langle[S_{+}^{\dag}({\bf q}),[H,S_{+}({\bf q})]]\rangle
\ee
with H given by eqn (1) on a square lattice.  Explicit evaluation of the double
commutator yields the result
\be
\int{\omega S^{\perp}({\bf q},\omega) d\omega} = 
\sum_{\bf k}(\varepsilon_{{\bf k}-{\bf q}} - \varepsilon_{\bf k})\langle
n_{{\bf k}\uparrow} - n_{{\bf k}\downarrow}\rangle 
+ 2h\langle S_{z}(-{\bf g})\rangle
\ee

where $ \varepsilon_{\bf k} = \Sigma_{i}te^{i{\bf k} \cdot {\bf R_{i}}} $.  Also
, it is easy to show [6] that
\be
t\sum_{i}(1-e^{i{\bf q} \cdot {\bf R_{i}}}) = t\sum_{i}(1-cos({\bf q} \cdot {\bf R_{i}}))
\ee
Using this in eqn (9) gives
\be
\langle[S_{+}^{\dag}({\bf q}),[H,S_{+}({\bf q})]]\rangle = \sum_{i}[q^{2}R_{i}^{2}t({\bf R_{i}})P({\bf R_{i}}) + 2hS_{0z}(-{\bf g})]
\ee
where $ P({\bf R_{i}}) = (1/2)\Sigma_{\bf k} e^{-i{\bf k} \cdot {\bf
R_{i}}}\langle n_{{\bf k}\uparrow} - n_{{\bf k}\downarrow}\rangle $.  
The $ t({\bf R_{i}})$ are matrix elements of the one-electron operators
between Wannier states, which fall off rapidly with $ R_{i} $.  So $ \Sigma_{i}(1/2)R_{i}^{2}t({\bf R_{i}})P({\bf R_{i}}) = L $
is well defined.  Also, $ S_{0z}(-{\bf g}) = (1/N)\langle S_{z}(-{\bf g}) \rangle $.  With these substitutions, eqn (11) becomes
\be
(1/2N)\langle[S_{+}^{\dag}({\bf q}),[H,S_{+}({\bf q})]]\rangle = (1/2)[Lq^{2} + 2hS_{0z}(-{\bf g})]
\ee
Substituting into eqn (5) yields
\be
S^{\perp}({\bf q}) = \int{S^{\perp}({\bf q},\omega) d\omega}
        \le \sqrt{[\frac{\chi^{\perp}}{2}({\bf q})(Lq^{2} + 2hS_{0z}(-{\bf g}))]}
\ee
Using the definition of $ S^{\perp}({\bf g}) $, we have [5]
\be
\Sigma_{\bf q} S^{\perp}({\bf q}+{\bf g}) \le (1/2)(N^{2}/2N) = N/4
\ee
Hence,
\be
(N/4) \ge s_{\bf g}^{2}\sum_{\bf q}{1\over[(1/2)\chi^{\perp}({\bf q})(Lq^{2}+2hS
_{0z}(-{\bf g}))]^{1/2}}
\ee
or
\be
 [S_{0z}(-{\bf g})]^{2} \le (N/4)\left[\sum_{\bf q}{1\over(\chi^{\perp}/2)[Lq^{2}+2hS_{0z}(-{\bf g})]^{1/2}}\right]^{-1}
\ee
Here, we used the fact that the perpendicular susceptibility $ \chi^{\perp}({\bf
 q}) $ approaches a constant value for small $ q $[5].
Replacing the sum over $ {\bf q} $ by an integral $\Sigma_{\bf q} \rightarrow \int{d^{D}q\over(2{\pi})^{D}}$, 
and letting $ q_{0} $ be the distance of the nea
rest Bragg reflection plane from the origin in q-space, we evaluate the integral
 to get
$I(1)=ln\left[(1/{\alpha})^{1/2} + (1+(1/{\alpha}))^{1/2}\right]$ in $D=1$,
while in $D=2$, we get
$I(2)=\left[q_{0}(1+{\alpha})^{1/2} - {\alpha}^{1/2}\right]$.
Here $ \alpha $ = $ (2hS_{0z}(-{\bf g}))\over(q_{0}^{2}L) $,
so that for small h, we get
\be
[S_{0z}(-{\bf g})]^{2} \le  (\frac{N\chi}{8I(1)})  \rightarrow 0 \hspace{1cm}
\mbox{as h $\rightarrow$ 0 in d=1}
\ee

\be
[S_{0z}(-{\bf g})]^{2} \le  (\frac{N\chi}{8I(2)}) > 0 \hspace{1cm}
\mbox{as h $\rightarrow$ 0 in d=2}
\ee
Hence, our analysis shows explicitly the absence of magnetic order in one dimension,
but yields yields only an upper bound at $ T=0 $ in 2d.  It is believed in
the literature that the 2d Hubbard model at $ T=0 $ exhibits antiferromagnetic long range order
 [7(a,b),8].  Together with the rigorous result on the spin-spin correlation
function of the Hubbard model on bipartite lattices [11], our result uses only
exact results and symmetries of the one-band Hubbard Hamiltonian on bipartite lattices to show that the
existence of antiferromagnetic LRO is not excluded in 2 dimensions.  We
emphasize that the derivation above does not rely on the applicability of the semiclassical
spin wave approximation, but treats the dominant $ T=0 $ quantum fluctuations.  We also notice
 that $ S_{0z}(-{\bf g},h) $ is a $\it {smooth} $ function of h for small h.
   We shall make use of this fact in what follows.

We now use the unitary particle-hole transformation for a bipartite lattice [9],
$ C_{i\uparrow} \rightarrow C_{i\uparrow} $,
$ C_{i\downarrow} \rightarrow \epsilon(i)C_{i\downarrow}^{\dag} $
with $ \epsilon(i)=1 $, i in A-sublattice, and equals $ -1$ otherwise.  The
Hubbard Hamiltonian is transformed to
\be
H = -t\sum_{ij\sigma}(C_{i\sigma}^{\dag}C_{j\sigma}+h.c) - U\sum_{i}n_{i\uparrow}n_{i\downarrow}
- ((h-U)/2)\sum_{i\sigma}n_{i\sigma}
\ee
and

\be
H_{ext}= - (\mu - (U/2))\sum_{i}(n_{i\uparrow}-n_{i\downarrow})
\ee
where $ H_{ext} $ is the part of the transformed hamiltonian that looks like a 
Zeeman term describing the linear coupling of the fermions to a magnetic field ${\it h} = (\mu -(U/2)) $.

In the case of a singlet ground state, $ 2\langle S_{i}^{+}S_{j}^{-}\rangle $ =
$ \langle S_{i}^{z}S_{j}^{z}\rangle $ = $ \varepsilon(i)\varepsilon(j)C_{ij} $ w
ith $ C_{ij} \ge 0$[11].  Under these conditions, $ |S_{i}^{z}|  \le   |S^{z}(-{
\bf g},h)| $.  Using this inequality,
the eqn for $ |[S_{0z}(-{\bf g},h)] | $ takes the form
\be
1-C[ln{\beta}]^{-1/2} \le \rho({\mu^{\prime}}) \le
1+C[ln{\beta}]^{-1/2}
\ee
where $ \beta = 2\left(q_{0}^{2}L/2(2\mu^{\prime}-U^{\prime})\right) $ in $d=1$.  
Also,
\be
1-C^{\prime}[(2/\beta)+\zeta]^{-1/2} \le \rho({\mu^{\prime}}) \le
1+C^{\prime}[(2/\beta)+\zeta]^{-1/2}
\ee
where $ \zeta > 0 $, in $ d=2 $.  Here $ C $ and $ C^{\prime} $ are constants, a
nd the eqs (23) and (24) are valid for small $ (2\mu^{\prime}-U^{\prime})$. Also, 
$ \rho({\mu}) = M^{-1}\Sigma_{i\sigma}\langle n_{i\sigma}\rangle $, the elect
ronic density per
site.  The above equations show that in one and two dimensions, for $ {\it small
} $ dopings, i.e for ${\mu^{\prime}} $ close to $ U^{\prime}/2 $, the density,
$\rho({\mu})$, is a continuous function of $\mu$.  According to the criterion fo
r phase separation requiring a
discontinuous variation in density with chemical potential [10], we conclude that
 the one-band Hubbard model near half-filling does not exhibit the phenomenon of
 phase separation in either one or two dimensions at $ T=0 $.  At exactly half-
filling, the ground state of the Hubbard model is expected to be insulating both
 in one- as well as in two dimensions for $ {\it any} $ $ U/t $ on a bipartite l
attice.  This holds for our case of a two-dimensional square lattice, and is
also the case for a one dimensional linear chain.
  At an arbitrary density away from the half-filled case, we define pseudospin operators
\be
J_{i}^{+}=\varepsilon(i)C_{i\uparrow}^{\dag}C_{i\downarrow}^{\dag},
J_{i}^{-}=[J_{i}^{+}]^{\dag},
J_{i}^{z}=(1-n_{i\uparrow}-n_{i\downarrow})/2
\ee
satisfying the usual pseudospin SU(2) algebra.  Without loss of generality, we
can set the magnetic field h to zero in what follows.  Choosing $ A = J^{+}({\bf
 q}+{\bf g}) $, and $ B = J^{+}({\bf q}) $, we can show, using the Pitaevskii
 and Stringari
inequality, that $ [J^{z}] $ is a smooth, continuous function of $ (\mu - (U/2))
 $ in one- and two dimensions at $ T=0 $.  We also reach the same conclusion by
using an argument due to Shen {\it et.al} [11], based  on the unitary mapping on
 bipartite lattices.
 The analysis carried out there shows that the positive U Hubbard model away from
 half-filling ($ S^{z} \ne 0 $) is equivalent to a negative U Hubbard model with
 non-zero $ J^{z} $.  Since the above derivation holds independent of the sign
of U, it
follows that $ [J^{z}] $ is a smooth, continuous function of $ (\mu - (U/2)) $,
as remarked earlier.This implies that the density is a continuous function of the
 chemical potential at {\it any} filling, and hence the criterion for phase separation leads
one to conclude that the one-band Hubbard model does not support phase separation
 at any filling in one- and two dimensions at $ T=0 $.
The above conclusions are likely to be unchanged by addition of a two-body term
describing nearest neighbor interactions at half-filling.  The situation away
from half-filling requires more care, because the n.n two-body term  breaks the pseudospin SU(2)
invariance away from n=1.  However, addition of terms describing next nearest 
neighbor (n.n.n) hopping in the Hubbard model will break particle-hole symmetry at
 half-filling, and thus invalidate the conclusions obtained above.  In addition,
 coupling to
lattice vibrations via a local Holstein-like coupling might affect our
conclusions.  Possible extensions of the work presented here to include these interesting
 cases is an open and interesting problem, and is left for the future.

Alongwith the demonstration of the impossibility of phase separation in the one-
band Hubbard model at finite T in one- and two dimensions [4], our demonstration
 of the impossibility of phase separation in the lower dimensional one-band Hubbard model has
interesting consequences {\it if} phase separation is intimately related to the
physics giving rise to high-$T_{c}$.
It has been suggested that phase separation (PS) in cuprate superconductors is a
 relevant issue [12,13,14] and may favor superconductive pair formation near the
 insulator-metal phase boundary.  We could say that the above derivation implies
 that it is
inappropriate to use the one-band Hubbard model to address the issue of phase 
separation in cuprate SCs.  This argument might not apply if the Hubbard
Hamiltonian is augmented with special terms ${\it e.g}$, n.n.n hopping terms.  In this 
case, particle-hole symmetry is inoperative, and precludes the use of formal
mathematical arguments leading to EPS.  The same argument holds for the extended
 Hubbard model (with nearest neighbor interactions) away from half-filling, where
$SU(2)$ symmetry is broken.  It should also be remarked
that the phase separation may probably be of electronic origin, rather than one
driven by electron lattice coupling [15].  Evidence of charge ordered phases [15
] near the PS phase could be interpreted as arising from coupling of lattice 
vibrations to a phase separated
 phase which has its origin in a purely electronic mechanism.

 On the other hand, phase separation arising from a purely electronic mechanism
 has a natural explanation in models with degenerate ground states.  To be
specific, the three band Hubbard model [16,17,18] (references in [1])
or the $t-t'-U$ Hubbard model [19] with/without phonons do show
numerical signatures of phase separation.  Both models break particle-hole
symmetry present in the simplest one-band Hubbard model, rendering the above
proof invalid.

  Phenomena akin to EPS have also been observed on mesoscopic length scales in
the colossal magnetoresistance (CMR) manganites [20], and discussed theoretically
 in terms of effective models.  It has been argued that the modification
of these mesoscale, phase separated regions by external magnetic fields holds
the key to the spectacular sensitivity of the transport properties of CMR
materials to small perturbations.  Interestingly, we see that manganites require
 a description in terms of multi-orbital Hubbard models (with ordered/disordered
 Jahn-Teller distortions).
The multi-orbital
Hamiltonian relevant for describing manganites is {\it not} $SU(2)$ invariant
in the orbital pseudospin sector, but has an effective $Z_{2}$ symmetry.
Thus, we expect a generalised Falicov-Kimball type of model to describe such
materials [21].  Hence, EPS found in manganites is in conformity with our result
 here, to the extent that the realistic model used for describing their physical
 behavior is not $SU(2)$ invariant from the outset.  Indeed, Freericks 
{\it et al.} [22] have {\it rigorously} proved EPS for the Falicov-Kimball
model for any dimension.  Ref.[21] finds
 a ground state degeneracy (within $d=\infty$) resulting from the mapping onto a
 Falicov-Kimball (electronic version of a binary alloy) model.
This would imply a jump in $n(\mu)$ as a function of
$\mu$, and EPS as a consequence.  In fact, such a model exhibits the electronic
analogue of {\it alloy} phase separation, extensively studied in context of
alloy physics [23].  Finally, this also implies that multi-orbital Hubbard models
 should generically exhibit electronic phase
separation: given that insulating, orbital ordered (OO) ground states in these 
models spontaneously break the Z$_{N}$ symmetry of the Hamiltonian 
($N$ is the number of {\it degenerate} orbitals), doping such models may
induce EPS close to the correlation driven,
first-order Mott transitions accompanied by melting of the OO insulating states.
  In the CMR case, this is manifested as
mesoscopic hole-rich (Jahn-Teller (JT) undistorted) clusters in the background of
 JT-distorted, hole-poor regions.  Their size
increases with modest external magnetic fields because the double-exchange induced
 increase in carrier kinetic energy overcomes
short-range OO, and this has indeed been proposed as a mechanism of CMR [20].

In conclusion, we have shown that the one-band Hubbard model does not show magnetic
 LRO in 1d, but our calculation yields only an upper bound on the sublattice
magnetization in 2d.  Hence, we conclude that AF LRO is not excluded in 2d at $T=0$.
  We have
  used the fact that the sublattice magnetization is a smooth function of the
applied magnetic field, along with a particle-hole transformation, to show that
 the Hubbard model does not support   phase separation at any filling at
$T=0$ in  one- and two
dimensions.  Extensions of the derivation to consider the effect of special
terms like those describing n.n.n hopping and coupling to phonons
(where EPS might
indeed be realised) are being studied and will be reported in future.

{\bf Acknowledgements:-}  I thank Prof. P. Fulde for his support and
encouragement, and MPIPKS, Dresden, for financial support.

\vspace{2cm}

 {\bf References.}

\vspace{1cm}
(1) See for e.g, "Phase Separation in Cuprate Superconductors", eds E. Sigmund
 and K. A. Muller (Springer-Verlag, Berlin, 1994).

(2) J. D. Jorgensten ${\it et.al}$, Phys. Rev. B ${\bf 38}$, 11337 (1988).

(3) C. Au ${\it et.al}$, unpublished.

(4) G. Su, Phys. Rev. B {\bf 54}, R8281 (1996).

(5) L. Pitaevskii and S. Stringari, J. Low. Temp. Phys. ${\bf 85}$, 377 (1991).

(6) N. D. Mermin and H. Wagner, Phys. Rev. Lett. ${\bf 17}$, 1133 (1966).

(7) E. H. Lieb and F. Y. Wu, Phys. Rev. Lett. $ {\bf 20} $, 1445 (1968).

(8) J. D. Reger and A. P. Young, Phys. Rev. B ${\bf 37}$, 5978 (1988).  Also, se
e S. Liang, Phys. Rev. B ${\bf 42}$ 6555 (1990).

(9) H. Shiba, Prog. Theoretical Physics. ${\bf 48}$, 2171 (1972).

(10) E. Dagotto, Rev. Mod. Phys. ${\bf 66}$, 763 (1994), and references therein.

(11) S. Q. Shen ${\it et. al}$, Phys. Rev. Lett. ${\bf 71}$, 25, 4238 (1993).

(12) See for e.g, articles in Ref. (1).

(13) M. Grilli, ${\it et. al}$, Phys. Rev. B ${\bf 45}$, 10805 (1992).

(14) M. Grilli, ${\it et. al}$, Phys. Rev. Lett. ${\bf 67}$, 259 (1991).

(15) P. C. Hammel, ${\it et. al}$, Phys. Rev. Lett. ${\bf 71}$, 440 (1993).

(16) C. M. Varma, Int. J. Mod. Phys. B ${\bf 3}$, 2083 (1989), and references th
erein.

(17) C. Sire, ${\it et. al}$, Phys. Rev. Lett. ${\bf 72}$, 15, 2478 (1994).

(18) P. B. Littlewood and C. M. Varma, Phys. Rev. B ${\bf 46}$, 405 (1992).

(19) M. Aichhorn {\it et al.}, Phys. Rev. B {\bf 76}, 224509 (2007).

(20) see E. Dagotto in ``Physics of Colossal Magnetoresistance Manganites''
(Springer Verlag) and references therein.

(21) V. Ferrari {\it et al.}, cond-mat/9906131. see also, T. V. Ramakrishnan,
J. Phys.: Condens. Matter {\bf 19}, 125211, (2007), and references therein.

(22) J. K. Freericks {\it et al.}, Phys. Rev. Lett. {\bf 88}, 106401
(2002).

(23) D. de Fontaine, in {\it Solid State Physics}, Vol. 47, Eds. H. Ehrenreich a
nd D. Turnbull, Academic Press (1994).

\end{document}